\documentclass{article}
\usepackage[numbers]{natbib}
 \usepackage[preprint]{neurips_2025}

\usepackage[utf8]{inputenc} 
\usepackage[T1]{fontenc}    
\usepackage{hyperref}       
\usepackage{url}            
\usepackage{booktabs}       
\usepackage{amsfonts}       
\usepackage{nicefrac}       
\usepackage{microtype}      
\usepackage{xcolor}         

\title{Dynamic ReAct: Scalable Tool Selection for Large-Scale MCP Environments
}

\author{%
  Nishant Gaurav \\
  \texttt{nishant@agentr.dev} \\
  \And
  Adit Akarsh \\
  \texttt{adit@agentr.dev} \\
  \AND
  Ankit Ranjan \\
  \texttt{ankit@agentr.dev} \\
  \And
  Manoj Bajaj \\
  \texttt{manoj@agentr.dev} \\
}

\begin{document}

\maketitle

\begin{abstract}
We present Dynamic ReAct, a novel approach for enabling ReAct agents to efficiently operate with extensive Model Control Protocol (MCP)\cite{mcp_community} tool sets that exceed the contextual memory limitations of large language models. Our approach addresses the fundamental challenge of tool selection in environments containing hundreds or thousands of available tools, where loading all tools simultaneously is computationally infeasible. We propose and evaluate five distinct architectures that progressively refine the tool selection process, culminating in a search-and-load mechanism that achieves intelligent tool selection with minimal computational overhead. Our experimental results demonstrate that the proposed approach reduces tool loading by up to 50\% while maintaining task completion accuracy, advancing the path towards truly general-purpose AI agents capable of dynamically adapting to diverse task environments.

\end{abstract}

\section{Introduction}
Large-scale LLM-based agents increasingly rely on external tool integrations to transcend context limits and improve performance~\cite{llm_paradigms_survey,tool_use_survey2024}.  
Model Context Protocols (MCP) are emerging as a unifying standard for managing diverse tool sets in these agents~\cite{mcp_community,anthropic_mcp}.
The proliferation of tools and APIs available to LLM-based agents presents a fundamental challenge: while the breadth of available tools continues to expand, the contextual memory of LLMs remains finite. Traditional approaches that load all available tools into an agent's context become impractical as tool registries grow beyond hundreds or thousands of entries. Recent surveys highlight key challenges facing LLM-based agents as tool registries expand, motivating scalable dynamic selection approaches~\cite{llm_paradigms_survey, scalemcp}. These limitations necessitate a dynamic approach to tool management that can intelligently select and load only the relevant subset of tools for a given task. 

We propose a Dynamic MCP ReAct Agent architecture that addresses this challenge through systematic tool retrieval and loading. Our approach leverages meta-tools—tools specifically designed to manage other tools—combined with semantic search capabilities to create an AutoAgent that adapts its tool set based on user requirements. The need for robust retrieval and scalable tool management is further underscored in recent work on dynamic model control protocols~\cite{scalemcp, mcp_zero}. This paper presents multiple architectural designs, evaluates their effectiveness, and identifies optimal configurations for production deployment.

\section{System Architecture}
\subsection{Core Components}
The Dynamic MCP ReAct Agent environment consists of four primary components that work in concert to enable intelligent tool selection and execution:
\begin{itemize}
    \item \textbf{LLM Client}: The central reasoning engine equipped with a lightweight system prompt that describes the operational framework without overwhelming the model with excessive instructions. This design choice ensures consistent instruction following while maintaining compatibility with third-party MCP clients.
    \item \textbf{Meta Tools}: A fixed set of tools permanently available to the LLM that facilitate the discovery, selection, and loading of task-specific MCP tools. These tools are characterized by their names, parameters, descriptions, and output formats, serving as the primary control mechanism for dynamic tool management.
    \item \textbf{Tool Registry}: A comprehensive repository containing all available MCP tools across various applications and domains. Each tool entry includes detailed descriptions optimized for semantic search and retrieval. While our experiments utilize a proprietary registry, the architecture supports integration with third-party tool repositories.
    \item \textbf{Vector Database}: A semantic search system that indexes tool names and descriptions, enabling rapid retrieval of relevant tools based on natural language queries. The database returns the top-k semantically similar tools for any given query, forming the foundation of our dynamic loading mechanism.
\end{itemize}

From these components, we identify seven distinct levers that govern the behavior of the system:

\begin{itemize}
    \item LLM Client (1 lever): System Prompt
    \item Meta Tools (4 levers): Names, Parameters, Descriptions, Output Format
    \item Tool Registry (1 lever): Tool Descriptions
    \item Vector DB (1 lever): Retrieval design that determines how effectively relevant tools are surfaced
\end{itemize}

\subsection{Design Principles}

Our design is guided by practical constraints and scalability considerations:

\begin{itemize}
    \item \textbf{System Prompt:} We keep the system prompt lightweight because LLMs often stop following instructions when overloaded. Furthermore, minimizing reliance on the prompt allows seamless integration with third-party MCP clients that typically do not permit system prompt modification.

    \item \textbf{Tool Registry:} The registry may originate from third parties. For this study, we employ our own tool registry, enabling us to experiment with and provide insights into writing richer tool descriptions that improve retrieval performance.

    \item \textbf{Control Assumptions:} In practice, the LLM Client and Tool Registry are often outside of our control. Consequently, our architectural exploration begins with Meta Tools as the primary lever for AutoAgent design. The Vector Database forms the second major lever under our control, where we evaluate multiple design strategies for retrieval accuracy and efficiency.
\end{itemize}

\section{Architectural Evolution}

\subsection{Baseline Architecture: Direct Semantic Search}

\textbf{Design.}  
In our initial design, the environment contains only three components:
\begin{itemize}
    \item The LLM client
    \item The full MCP tool set
    \item A vector database indexing the tool registry
\end{itemize}

When a user sends a message, it is passed directly to the vector database as a query. The database retrieves the top-$k$ semantically similar tools, which are then bound to the LLM to form a ReAct agent. The agent is finally invoked with the original user message. Such baseline approaches relying purely on semantic vector search have been shown to suffer from retrieval imprecision and context saturation in large-scale environments~\cite{graphragmcp, mcp_zero}, but we have the results below for completeness.

\textbf{Advantages.}
\begin{itemize}
    \item No additional tool calls are needed before loading MCP tools.
    \item The pipeline remains simple: User query $\rightarrow$ Vector search $\rightarrow$ Tool binding $\rightarrow$ Execution.
\end{itemize}

\textbf{Issues with Semantic Search.}

\textit{Lack of specificity.} Even with a well-phrased query, the most relevant tool may not rank at the top. To avoid missing it, a large $k$ must be used, which floods the LLM’s context with unnecessary tools.

\textbf{Example:} Query $\rightarrow$ “Create a list of unsubscribe links for mails from my inbox”

\renewcommand{\arraystretch}{1.5}
\begin{table}[h]
\caption{Retrieved tools for unsubscribe query under baseline semantic search.}
\centering
\begin{tabular}{p{0.45\linewidth} p{0.5\linewidth}}
\hline
\textbf{Retrieved Tools (subset)} & \textbf{Description} \\
\hline
\texttt{mailchimp\_\_reports\_list..} & Retrieves a list of unsubscribed members for a specified campaign. \\
\texttt{braze\_\_list\_unsubscribes} & Query list of unsubscribed email addresses. \\
\texttt{mailchimp\_\_reports\_list..} & Retrieves a list of subscribers who clicked a specific link in a campaign. \\
\texttt{mailchimp\_\_reports\_get..} & Retrieves information about an unsubscribed member in a campaign report. \\
\texttt{mailchimp\_\_lists\_get\_all..} & Retrieves all abuse reports for a specific list. \\
\texttt{outlook\_\_list\_user\_messages} & Retrieves messages from a user’s mailbox with filtering and search. \\
\hline
\end{tabular}
\end{table}

Problem: While some tools mention “unsubscribe,” none directly solve the intended task of extracting unsubscribe links from emails. The results are noisy, drawn from different providers, and require a large $k$ to catch even tangentially relevant tools.

\textit{Mixed results for complex queries.} When queries require multiple coordinated actions across applications, the system often retrieves only tools from a single domain, missing the multi-application requirement.

\textbf{Example:} Query $\rightarrow$ “Monitor my Twitter mentions and DMs from the past 48 hours and create a response priority list in Google Sheets.”

\renewcommand{\arraystretch}{1.5}
\begin{table}[h]
\caption{Retrieved tools for multi-step Twitter $\rightarrow$ Google Sheets query.}
\centering
\begin{tabular}{p{0.45\linewidth} p{0.5\linewidth}}
\hline
\textbf{Retrieved Tools (subset)} & \textbf{Description} \\
\hline
\texttt{twitter\_\_get\_user\_mentions} & Retrieve mentions of a Twitter user. \\
\texttt{twitter\_\_get\_tweets\_firehose\_stream} & Retrieve tweets in real-time from the Twitter firehose. \\
\texttt{twitter\_\_get\_user\_tweet\_timeline} & Retrieve tweets posted by a specific Twitter user. \\
\texttt{twitter\_\_get\_list\_tweets} & Retrieve tweets from a Twitter list. \\
\texttt{twitter\_\_get\_usage\_tweets} & Retrieve usage statistics of tweets. \\
\hline
\end{tabular}
\end{table}

Problem: The retrieved set is entirely Twitter-focused, with no tools for Google Sheets. As a result, the system cannot complete the multi-step workflow (monitor + prioritize + log in Sheets).

While this design offers simplicity and low overhead, its reliance on raw semantic search leads to overloaded or incomplete tool sets. These issues motivate the need for meta-tools and query refinement strategies in subsequent architectures.

\subsection{Meta-Tool for Intelligent Search Query Construction}

\textbf{Design.}  
In this design, the vector search is exposed to the LLM as a meta tool. Instead of directly sending the user’s message as a query, the LLM first constructs the search queries. Each such query is atomic (i.e., focused on performing a single action), which results in more relevant tool retrieval. The results of this search are then loaded into the LLM, forming a ReAct agent that proceeds to act on the user’s original message.

\textbf{Advantages.}
\begin{itemize}
    \item \textbf{Intelligent Queries:} Since the LLM constructs the search queries, they are more targeted and specific. The decomposition into atomic queries improves retrieval quality compared to the raw message approach.
\end{itemize}

\textbf{Example.} User Query: “Create a list of unsubscribe links for mails from my inbox.”  
(Note: The user has connected only Gmail, among Gmail, Outlook, and other mail aggregator applications. Hence the constructed query focuses specifically on Gmail.)

Constructed search query: “search Gmail emails by date and filter promotional emails.”

\renewcommand{\arraystretch}{1.5}
\begin{table}[h]
\caption{Retrieved tools using intelligent query construction.}
\centering
\begin{tabular}{p{0.45\linewidth} p{0.5\linewidth}}
\hline
\textbf{Retrieved Tools} & \textbf{Description} \\
\hline
\texttt{google-mail\_\_create\_filters} & Set up new Gmail filters with criteria and automation rules. \\
\texttt{google-mail\_\_delete\_filters} & Remove Gmail filters and their automation rules. \\
\texttt{google-mail\_\_list\_filters} & Retrieve all Gmail filters and their automation settings. \\
\texttt{google-mail\_\_get\_filters} & Fetch Gmail filter configurations by filter ID. \\
\texttt{google-mail\_\_list\_drafts} & Retrieve and format a list of email drafts from Gmail. \\
\texttt{google-calendar\_\_list\_events} & Retrieve events from the user’s calendar. \\
\texttt{hubspot\_\_list\_marketing\_emails} & Retrieve all marketing emails in HubSpot. \\
\texttt{google-mail\_\_list\_messages} & List messages from the user’s Gmail mailbox. \\
\hline
\end{tabular}
\end{table}

\textbf{Disadvantages.}
\begin{itemize}
    \item Extra Overhead: An additional LLM call is required for constructing the search query whenever an MCP tool has to be executed.
    \item High Tool Count: The number of tools retrieved remains large, since vector search is still imprecise. The $k$ must be set high (at least 10 in our examples) to avoid missing relevant tools.
\end{itemize}

\subsection{Search and Load Architecture: Deliberate Tool Selection}

\textbf{Design.}  
In this design, two meta tools are bound to the LLM:

\begin{itemize}
    \item \texttt{search\_tools}: Takes in a list of string queries and performs a two-level search.
    \begin{itemize}
        \item For every query, it retrieves $k_1 = 20$ candidate tools.
        \item Repetitions are removed, and the number of tools per application is capped at $k_2 = 5$.
        \item The final deduplicated list of tools is returned.
    \end{itemize}
    \item \texttt{load\_tools}: After parsing the search results, the LLM selects specific tools and calls \texttt{load\_tools} with their IDs. This binds the selected tools to the LLM before invoking it with the user query.
\end{itemize}

\textbf{Advantages}
\begin{itemize}
    \item \textbf{Intelligent Queries:} Queries are constructed by the LLM and remain atomic, leading to better retrieval quality.
    \item \textbf{Multi-Query Design:} Multiple queries can be executed in a single search, reducing the number of LLM calls.
    \item \textbf{Deliberate Loading of Tools:} The LLM carefully selects tools from the retrieved results, typically loading fewer than 5. This contrasts with vector search–based designs, which require a high $k$ (10 or more) and load all results indiscriminately.
\end{itemize}

\textbf{Example} \\
User Query: \emph{“Monitor my Twitter mentions and DMs from the past 48 hours and create a response priority list in Google Sheets.”}

Constructed queries:
\begin{itemize}
    \item “retrieve Twitter mentions and direct messages”
    \item “create Google Sheets spreadsheet with data”
\end{itemize}

\begin{table}[h]
\caption{Loaded tools for Twitter + Google Sheets workflow.}
\centering
\begin{tabular}{p{0.45\linewidth} p{0.5\linewidth}}
\hline
\textbf{Retrieved Tools} & \textbf{Description} \\
\hline
\texttt{twitter\_\_get\_user\_mentions} & Retrieve mentions of a Twitter user \\[2mm]
\texttt{twitter\_\_get\_dm..} & Retrieve direct message events by participant ID \\[2mm]
\texttt{google\_sheet\_\_create\_spreadsheet} & Create a new spreadsheet in Google Sheets \\[2mm]
\texttt{google\_sheet\_\_write\_values\_to\_sheet} & Write values into a Google Sheets spreadsheet \\
\hline
\end{tabular}
\end{table}

\textbf{Disadvantages}
\begin{itemize}
    \item Overhead: A minimum of two extra calls (\texttt{search} + \texttt{load}) are required whenever the LLM needs to execute an MCP tool.
\item Irrelevant Application Results: Applications that match semantically but are not useful may still appear.
  Example: Query \emph{“search Gmail emails by date and filter promotional emails”} included tools from \texttt{google\_mail}, \texttt{resend}, \texttt{mailchimp}, \texttt{unipile}, \texttt{klaviyo}, and \texttt{scraper}.
\end{itemize}

\subsection{Application-Aware Architecture: Hierarchical Search}

\textbf{Design.}  
In this design, the LLM has access to three meta tools:

\begin{itemize}
    \item \texttt{search\_apps} – Retrieves applications relevant to a query.
    \item \texttt{search\_tools} – Searches for tools relevant to the query, optionally filtered by application.
    \item \texttt{load\_tools} – Loads the selected tools into the LLM before acting on the user’s request.
\end{itemize}

This allows the LLM to first identify available apps, then narrow the search to tools within the most relevant applications.

\textbf{Advantages}
\begin{itemize}
    \item \textbf{Application Awareness:} \texttt{search\_apps} enables the LLM to check available applications before searching for tools.
    \item \textbf{Focused Tool Retrieval:} By filtering searches to specific apps, irrelevant matches are reduced, improving tool selection accuracy.
\end{itemize}

\textbf{Example}  
User Query: \emph{“Create a list of unsubscribe links for mails from my inbox.”}

App search results: Resend, Mailchimp, Unipile, Klaviyo, LinkedIn Scraper, Firecrawl, Braze, SendGrid, Calendly, Google Mail, Microsoft Outlook, Apollo, Spotify, HTTP Tools, Fireflies.

Tool search results (with $k=5$ per app):

\begin{table}[h]
\caption{Example tool search results with app filtering.}
\centering
\begin{tabular}{p{0.45\linewidth} p{0.5\linewidth}}
\hline
\textbf{Retrieved Tools} & \textbf{Description} \\
\hline
\texttt{google\_mail\_\_list\_messages} & Retrieve messages from Gmail \\ 
\texttt{google\_mail\_\_get\_all\_filters} & Retrieve all Gmail filters \\ 
\texttt{google\_mail\_\_list\_labels} & List all Gmail labels \\ 
\texttt{google\_mail\_\_list\_drafts} & Retrieve Gmail drafts \\ 
\texttt{google\_mail\_\_delete\_filter} & Delete a Gmail filter \\ 
\texttt{outlook\_\_user\_delete\_message} & Permanently delete a message from Outlook \\ 
\texttt{outlook\_\_list\_user\_messages} & Retrieve a list of messages from Outlook \\ 
\texttt{outlook\_\_list\_message\_attachments} & Retrieve attachments from an Outlook message \\ 
\texttt{outlook\_\_get\_next\_page} & Retrieve the next page of Outlook messages \\ 
\texttt{outlook\_\_get\_current\_user\_profile} & Retrieve the current Outlook user’s profile \\
\hline
\end{tabular}
\end{table}

\textbf{Disadvantages}
\begin{itemize}
    \item \textbf{Extra Overhead:} The \texttt{search\_apps} step introduces an additional call without significantly improving accuracy.
    \item \textbf{Redundant Functionality:} Similar precision can be achieved by integrating application filtering directly inside \texttt{search\_tools}, reducing calls.
\end{itemize}

\subsection{Fixed Tool Set Architecture: Consistent Context}

\textbf{Design.}  
In this approach, the LLM operates with a fixed set of four meta tools. Retrieved MCP tools are not bound directly; instead, functionality is accessed dynamically.

Available meta tools:
\begin{itemize}
    \item \texttt{search\_app} – Retrieve applications relevant to a query.
    \item \texttt{search\_tool} – Search for tools, optionally filtering by application.
    \item \texttt{get\_tool\_info} – Takes in tool IDs and returns their docstring, input schema, and output schema.
    \item \texttt{call\_tool} – Executes a tool with a given ID and arguments, returning its output schema.
\end{itemize}

\textbf{Advantages}
\begin{itemize}
    \item \textbf{Caching Efficiency:} Since the bound tool set does not change, caching can be applied effectively, leading to cost savings.
\end{itemize}

\textbf{Disadvantages}
\begin{itemize}
    \item \textbf{Performance Degradation:} LLMs are optimized for directly bound tools; longer conversations suffer as the LLM fails to use \texttt{get\_tool\_info} and \texttt{call\_tool} efficiently.
    \item \textbf{Indirection Overhead:} Tool usage requires multiple steps (retrieving schema + invoking), adding friction compared to directly loaded tools.
\end{itemize}

\section{Implementation Optimizations}

\subsection{Search Function Design}
We experimented with multiple designs of the \texttt{search\_tools} meta tool, focusing on both input parameterization and the underlying retrieval strategy.

\subsubsection{App-Filtered vs. Query-Only Search}
We implemented two variants of the \texttt{search\_tools} function:

\begin{itemize}
    \item \textbf{App-filtered search:} Accepts a list of \{query, app\_id\} pairs, where the \texttt{app\_id} field is optional. This design allows the LLM to restrict retrieval to a specific application when such context is available.
    \item \textbf{Query-only search:} Accepts a list of queries (strings) without requiring an explicit application filter.
\end{itemize}

In practice, we observed that the query-only variant was preferred by the LLM in most cases, even when the app-filtered version was available. Interestingly, when the LLM did provide an app filter, the unfiltered search often retrieved the same tools by including the app name directly in the query string. This suggests that app-level filtering does not significantly improve retrieval performance, as the semantic vector search captures app context effectively when named explicitly in queries.

\subsubsection{LLM-Based Search Approach}
In addition to vector retrieval, we also implemented an LLM-based search method. In this design, a lightweight secondary LLM is initialized with a cached system prompt containing all tool descriptions. Since the system prompt remains fixed, caching lowers costs, while keeping this LLM separate preserves the contextual memory of the main agent.

This approach demonstrated strong accuracy for small tool sets but failed to scale for large registries. The main challenges were:
\begin{itemize}
    \item \textbf{Context limitations:} The fixed prompt quickly exceeded token limits when handling more than a few hundred tools.
    \item \textbf{Computational overhead:} Maintaining and querying this secondary LLM introduced significant latency and cost.
\end{itemize}

These scaling issues are consistent with prior findings~\cite{rag-mcp}. As a result, we adopted vector-based retrieval as our production strategy, which scales efficiently to thousands of tools while maintaining competitive accuracy.

\subsection{Output Formatting and Guidance}
Several refinements to output formatting significantly improved reliability:

\begin{itemize}
    \item \textbf{Reminder to load tools:} Adding an explicit reminder at the end of \texttt{search\_tools} output prevents premature execution attempts by the LLM on tools that have not yet been bound.
    \item \textbf{User-specific app connections:} Many applications overlap in functionality (e.g., Gmail vs. Outlook vs. Resend for sending emails). To resolve ambiguity, search results include information on which apps the user has connected. The LLM is instructed to prefer connected applications when breaking ties, avoiding unnecessary user queries.
    \item \textbf{System prompt variants:} We tested prompts with explicit lists of app names versus prompts without them. Including app names biased the agent towards loading tools from those apps unnecessarily. In our final implementation, we omitted explicit app lists to minimize this bias.
\end{itemize}

\textbf{Example:}
\begin{itemize}
    \item \textbf{Query:} Generate a comparison table of SaaS tools for project management, including pricing, features, and user ratings
    \item \textbf{Observed Behavior:} The agent loaded tools such as \texttt{airtable\_\_create\_record}, \texttt{coda\_\_list\_tables}, and \texttt{serpapi\_\_search} to build a comparison table. However, the user had not specified exporting results into a particular app. The redundant loading of Airtable and Coda illustrates the importance of controlling tool bias via prompt and formatting strategies.
\end{itemize}

\subsection{Default Tools Integration}
We observed that the agent frequently attempted to search for specialized tools to perform generic tasks (e.g., creating tables, conducting web searches). These futile searches often failed, despite the existence of common reusable functionality.

To mitigate this, we introduced default tools that are always available alongside meta tools:
\begin{itemize}
    \item \texttt{create\_table} – for tabular data generation
    \item \texttt{web\_search} – for general-purpose web queries
\end{itemize}

These defaults prevent wasted searches and improve performance on broad queries.

\textbf{Example:}
\begin{itemize}
    \item \textbf{User Query:} Find and summarize the key takeaways from the latest earnings calls of FAANG companies
    \item \textbf{Search Queries Generated:}
    \begin{itemize}
        \item earnings calls financial data
        \item company financial reports
        \item stock market earnings information
        \item financial market data stock prices
        \item SEC filings company reports
        \item news financial earnings
    \end{itemize}
    After adding the \texttt{web\_search} tool, instead of searching for multiple financial APIs, the agent leveraged the default \texttt{web\_search} tool to efficiently answer the query.
\end{itemize}

\section{Vector Retrieval Optimization}
Recent studies find that context-enriched embeddings substantially improve retrieval accuracy over baseline vector models~\cite{scalemcp, graphragmcp}. Hence, in addition to the above structural changes to the agent setup, we also perform experiments and modifications to the vector database.

\subsection{Experimental Setup}
We conducted extensive experiments to optimize our vector retrieval pipeline. The evaluation focused on precision (Top-5 accuracy) and recall (Top-10 accuracy) using a diverse test suite of tool-related queries.

\subsection{Embedding Model Performance}
Our baseline used OpenAI’s \texttt{text-embedding-3-large} model, achieving 40\% Top-5 and 64\% Top-10 accuracy. We then tested Voyage AI models:

\begin{itemize}
    \item \textbf{voyage-context-3:} 48\% Top-5, 68\% Top-10
    \item \textbf{voyage-3-large:} 56\% Top-5, 68\% Top-10
\end{itemize}

\subsection{Context Enrichment Strategy}
The most significant improvement came from programmatically enriching tool descriptions before embedding. Using Anthropic’s Sonnet 4, we generated additional context describing implicit functionalities and use cases not explicitly stated in the tool documentation.

\begin{itemize}
    \item \textbf{voyage-context-3 + Sonnet context enrichment:} 60\% Top-5, 68\% Top-10
\end{itemize}

This represented a 50\% relative improvement in Top-5 accuracy over the baseline.

\subsection{Hybrid Search Evaluation}
We also explored a hybrid search combining semantic search (vector-based) with BM25 keyword retrieval.

\begin{itemize}
    \item \textbf{voyage-context-3 + Sonnet context + BM25:} 56\% Top-5, 72\% Top-10
\end{itemize}

While recall improved, precision degraded slightly. BM25 occasionally promoted tools with keyword overlap but weak semantic relevance (e.g., including \texttt{Google Maps Search} in results for a general web search). This confirmed our decision to prioritize vector-only methods for production, where precision is paramount.

\subsection{Comparative Analysis of Queries}
\textbf{Query:} ``send email''

\begin{itemize}
    \item \textbf{Baseline (OpenAI embeddings):} Results dominated by unrelated Braze tools. \texttt{resend\_\_send\_email} ranked \#4, \texttt{google\_mail\_\_send\_email} ranked \#6, and \texttt{outlook\_\_send\_mail} was absent from the Top-10.
    \item \textbf{Optimized (Voyage + Context):} All three expected tools appeared in the Top-5:
    \begin{itemize}
        \item \texttt{outlook\_\_send\_mail} (\#1)
        \item \texttt{google\_mail\_\_send\_email} (\#2)
        \item \texttt{resend\_\_send\_email} (\#4)
    \end{itemize}
\end{itemize}

\textbf{Query:} ``web search general information''

\begin{itemize}
    \item \textbf{Optimized (Voyage + Context):} Perfect Top-5 retrieval of expected tools:
    \begin{itemize}
        \item \texttt{serpapi\_\_web\_search}, \texttt{firecrawl\_\_search}, \texttt{tavily\_\_search\_and\_summarize}, \texttt{perplexity\_\_answer\_with\_search}, \texttt{exa\_\_search\_with\_filters}
    \end{itemize}
    \item \textbf{Hybrid (Voyage + Context + BM25):} One irrelevant tool, \texttt{serpapi\_\_google\_maps\_search}, entered the Top-5 due to keyword overlap (``search''), replacing the semantically correct \texttt{exa\_\_search\_with\_filters}.
\end{itemize}

\subsection{Embedding Model Comparison}
\begin{table}[h!]
\caption{Comparison of embedding models and retrieval strategies.}
\centering
\begin{tabular}{|l|c|c|}
\hline
\textbf{Embedding Model} & \textbf{Top-5 (\%)} & \textbf{Top-10 (\%)} \\
\hline
OpenAI (\texttt{text-embedding-3-large}) & 40 & 64 \\
voyage-context-3 & 48 & 68 \\
voyage-context-3 + Sonnet context enrichment & 60 & 68 \\
voyage-context-3 + Sonnet context + BM25 & 56 & 72 \\
voyage-3-large & 56 & 68 \\
voyage-3-large + Sonnet context & 56 & 68 \\
\hline
\end{tabular}

\end{table}

\section{Discussion}

\subsection{Architecture Selection}
Our experiments reveal trade-offs between the architectural patterns explored.  

\begin{itemize}
    \item \textbf{Search and Load architecture:} Emerges as the optimal balance between precision and efficiency, requiring only two additional LLM calls while achieving high tool selection accuracy.
    \item \textbf{Fixed Tool Set architecture:} Shows promise for cost optimization through caching, but suffers performance degradation in longer conversations where agents tend to rely on direct tool calls.
\end{itemize}

\subsection{Scalability Considerations}
The Dynamic MCP ReAct Agent demonstrates strong scalability by loading only the tools necessary for a given query. This ensures constant memory usage regardless of registry size. Since our vector database supports efficient approximate nearest neighbor search, retrieval scales logarithmically with registry growth, making the system robust to thousands (or eventually millions) of tools.

\subsection{Limitations and Future Work}
Several open challenges remain:

\begin{itemize}
    \item \textbf{Tool documentation quality:} Current performance relies heavily on high-quality tool descriptions. Poorly documented tools significantly reduce retrieval accuracy. Automating the generation and validation of enriched descriptions could mitigate this issue.
    \item \textbf{Multi-tenant environments:} Our current architecture assumes a single user with a unified tool registry. Supporting multiple tenants with user-specific permissions and application connections requires additional architectural considerations.
    \item \textbf{Adaptive search parameters:} Dynamically adjusting $k$ values (number of results retrieved) based on query complexity could improve efficiency by avoiding over-retrieval for simple tasks while maintaining robustness for complex ones.
    \item \textbf{Learning from execution success:} Incorporating reinforcement learning from tool execution outcomes could enable the agent to improve its retrieval and selection strategies over time.
\end{itemize}
Many open issues, including multi-tenant support, adaptive retrieval, and leveraging execution feedback for learning, remain active areas of research~\cite{llm_paradigms_survey, graphragmcp}.

\section{Conclusion}
This paper presents a comprehensive solution to the challenge of managing extensive tool registries in LLM-based agents through dynamic tool selection and loading. Our Dynamic MCP ReAct Agent architecture, combining intelligent meta-tools with optimized vector retrieval, achieves significant improvements in both efficiency and accuracy compared to static tool loading approaches.

The key contributions of this work include:

\begin{enumerate}
    \item A systematic evaluation of five different architectural patterns for dynamic tool management.
    \item Demonstration that context-enriched embeddings improve tool retrieval accuracy by 50\%.
    \item Practical implementation strategies that balance performance with computational efficiency.
\end{enumerate}

Our Search and Load architecture with \texttt{voyage-context-3} embeddings and Sonnet-generated context enrichment provides a production-ready solution that scales to thousands of tools while maintaining high accuracy in tool selection.

As the ecosystem of available tools and APIs continues to expand, dynamic tool management becomes increasingly critical for practical agent deployment. The architectures and optimizations presented in this paper provide a foundation for building scalable, efficient agents capable of adapting to diverse user requirements without overwhelming system resources. The modular design ensures compatibility with existing MCP infrastructure while enabling continuous improvements in retrieval accuracy and execution efficiency.

\bibliography{main}

\end{document}